\documentclass[nofootinbib,preprint,amsmath,amssymb]{revtex4}

\usepackage{graphicx}
\usepackage{dcolumn}
\usepackage{bm}
\newcommand{\be}{\begin{equation}}
\newcommand{\ee}{\end{equation}}
\newcommand{\bea}{\begin{eqnarray}}
\newcommand{\eea}{\end{eqnarray}}

\begin{document}
\title{Unconventional cosmology on the (thick) brane}

\author{Ignacio Navarro$^1$}
\author{Jos\'e Santiago$^2$}

\affiliation{ ~$^1$IPPP, Department of Physics, University of
Durham, DH1 3LE Durham, UK \\~$^2$Fermi National Accelerator
Laboratory, P.O. Box 500,
  Batavia, IL 60510, USA }

\begin{abstract}

We consider the cosmology of a thick codimension 1 brane. We
obtain the matching conditions leading to the cosmological
evolution equations and show that when one includes matter with a
pressure component along the extra dimension in the brane
energy-momentum tensor, the cosmology is of non-standard type. In
particular one can get acceleration when a dust of
non-relativistic matter particles is the only source for the
(modified) Friedman equation. Our equations would seem to violate
the conservation of energy-momentum from a 4D perspective, but in
5D the energy-momentum is conserved. One could write down an
effective conserved 4D energy-momentum tensor attaching a ``dark
energy'' component to the energy-momentum tensor of matter that
has pressure along the extra dimension. This extra component could,
on a cosmological scale, be interpreted as matter-coupled
quintessence. We comment on the effective 4D description of this effect in terms of the time evolution of a scalar field (the 5D radion) coupled to this kind of matter.

\end{abstract}

\preprint{IPPP/05/19, DCPT/05/38} 
\preprint{FERMILAB-PUB-05-195-T}
\maketitle

The quest for a modified theory of gravity has taken renewed
impetus with the recent cosmological observations implying that
the expansion of our universe is accelerating \cite{Riess:1998cb}.
And given the technical and conceptual difficulties that
modifications of 4D General Relativity present the braneworld idea
is also a natural arena in which to look for non-standard behavior
of gravitational dynamics. The best studied case is the
codimension one braneworld, where one can find 4D gravity on the
brane even in the presence of an infinite extra dimension, when
the curvature of the bulk spacetime is negative and its volume
finite \cite{Randall:1999vf}. Usually the brane is modeled as a
distributional source in the energy-momentum tensor (EMT) of zero
thickness, and in this case the cosmology has been obtained and
analyzed in detail \cite{Binetruy:1999ut,Ida:1999ui}.
Unfortunately the non-standard features of the braneworld
cosmology (the presence of a ``dark radiation'' term, to be
identified with the mass of a bulk black hole \cite{Ida:1999ui})
did not shed any light on the apparently bizarre energy
composition of our universe. In this letter we show that when one
drops the infinitesimally thin idealization in the modeling of
the brane, and one includes the possibility of having a pressure
component in the brane EMT along the extra dimension one gets
non-standard gravitational dynamics.

Since we are interested in the cosmological behavior of the
braneworld we take the metric ansatz
\be ds^2 = n^2(r,t)dt^2 - a^2(r,t)d{\bf x}^2
- e^{2\phi(r,t)}dr^2,
\ee
where the (only) brane
will be now an extended object filling the region with $|r|<
\epsilon$ and we have taken flat spatial sections for simplicity.
Furthermore we assume a $Z_2$ symmetry with respect to $r=0$. Now
we can obtain the matching conditions, that make reference to the
metric and its first derivatives evaluated at $r=\epsilon$, by
integrating the $\mu\nu$ Einstein equations 
(in the form $M_\ast^3 R^\mu_\nu=T^\mu_\nu-\delta^\mu_\nu T/3$) 
in the $|r|<\epsilon$ region giving
\bea
e^{-\phi}\frac{2 n^\prime}{n}|_\epsilon = \frac{1}{M_\ast^3}
\left[\frac{2}{3}\rho + p
+ \frac{1}{3}p_r \right],\label{matching1}
\\
e^{-\phi}\frac{2 a^\prime}{a}|_\epsilon =
\frac{1}{M_\ast^3} \left[\frac{1}{3}\left(p_r-\rho\right)\right],
\label{matching2}
\eea where a prime denotes differentiation with respect to $r$,
$M_\ast$ is the five-dimensional reduced Planck mass and the
energy densities $\rho$, $p$ and $p_r$ are defined as the
integration of the corresponding elements of the full 5D EMT from
$r=-\epsilon$ to $r=\epsilon$:
\bea
\rho \equiv
\frac{1}{na^3|_{\epsilon}}\int_{-\epsilon}^\epsilon T_0^{0} n a^3
e^\phi dr, \\
p\equiv
-\frac{1}{na^3|_{\epsilon}}\int_{-\epsilon}^\epsilon T_x^{x} n a^3 e^\phi
dr,
\\
p_r\equiv -\frac{1}{na^3|_{\epsilon}} \int_{-\epsilon}^\epsilon
T_r^{r} n a^3 e^\phi dr.
\eea 
Once one considers a thick brane these
matching conditions are only approximate, as we have neglected terms
in the integration of the Einstein equations that are proportional
to the derivatives of the metric along the parallel brane
coordinates. It can be seen that, provided the brane is
\textit{thin} -- in a sense that we will define more precisely
below, see eq.(\ref{approximations})-- these terms are indeed
negligible for the cosmological solutions we will consider (in
fact they are negligible in most situations, see
\cite{Navarro:2004di} for a discussion of matching conditions for
thick codimension 1 and 2 branes). Notice that we have considered
a non-zero $T_{r}^{r}$ inside the brane, a component of the brane
EMT that is usually taken to be zero in the thin brane limit. 
However, for a brane with finite width this quantity cannot be
assumed to be zero or even time-independent in general, as can be
seen from the equation of conservation of the EMT whose $r$
component reads
\be
\left(T_r^r\right)^\prime + \left(
\frac{n^\prime}{n}
+
\frac{3 a^\prime}{a}
\right)T_r^r
-\frac{n^\prime}{n}T_0^0
-\frac{3 a^\prime}{a}T_x^x
=0.\ee We see that when
$n^\prime$ and $a^\prime$ are not zero \textit{inside} the brane
we also need in general a nontrivial $r$-dependence of $T_r^r$.

We will also assume that in the bulk ($|r|>\epsilon$) the EMT is
just that of a cosmological constant, $T^M_{N}=\delta^M_N\Lambda$.
With these assumptions we are ready to obtain the equations
governing the cosmological evolution of our braneworld. As in the
$\delta$-like case we just have to evaluate the $rr$ and $0r$
components of the full 5D Einstein equations at $r=\epsilon$.
These are, respectively 
\bea
-3 \mathrm{e}^ {-2\phi}\frac{a^\prime}{a} \left(
\frac{a^\prime}{a}+n^\prime \right)+
3\left(\frac{\ddot{a}}{a}+\frac{\dot{a}^2}{a^2}\right)
&=& \frac{\Lambda}{M_\ast^3}, \\
\partial_0 \left[-6 e^{-\phi} \frac{a^\prime}{a} \right] + 6
\frac{\dot{a}}{a} e^{-\phi}\left[
n^\prime-\frac{a^\prime}{a} \right]&=&0,
\eea
where a dot denotes differentiation
with respect to $t$, we have taken $n(\epsilon,t)=1$ 
and all the functions in here and in the
following are evaluated at $r=\epsilon$ unless stated otherwise.
We consider now a splitting of the brane EMT in a constant,
time independent background part (the so-called brane tension) and
a time-dependent matter contribution ($\rho_m$) with an arbitrary
(but assumed constant) equation of state, so taking
\be
\rho = T +
\rho_m, \quad 
p = -T + \omega \rho_m,\quad 
p_r = -T_r +\omega_r \rho_m,
\ee
and using our matching conditions,
eqs.(\ref{matching1},\ref{matching2}),
we can expand our equations at first order in
$\rho_m$ to get
\bea
3\left(
\frac{\ddot{a}}{a}+\frac{\dot{a}^2}{a^2}\right) &=&
\frac{1}{12 M_\ast^6} \Big[ 2(T+T_r)^2+(T+T_r)(1-3w-4w_r)\rho_m\Big]
+\frac{\Lambda}{M_\ast^3},
\label{friedtr}
\eea
\be
\dot{\rho}_m + 3\frac{\dot{a}}{a}\left(1 + \omega \right)\rho_m -
\omega_r \dot{\rho}_m=0.\label{EMTcons}
\ee
It is immediate to see that if $\omega_r=0$ we
recover a conventional cosmology (with a cosmological constant
given in Planck mass units by $\Lambda_{eff}=\frac{1}{6M_\ast^6}
\left(T + T_r\right)^2 +\frac{\Lambda}{M_\ast^3}$) up to ${\cal O}(\rho_m^2)$
terms as for the
infinitesimally thin brane. The only difference with the well
studied $\delta$-like brane would be an extra contribution to the
inverse Planck mass and cosmological constant that can be taken
into account by shifting $T \rightarrow T + T_r$. But for a
generic value of $\omega_r$ these equations look quite different
from the conventional cosmological equations obtained in 4D
General Relativity. Notice that in particular the second of these
equations would seem to violate the 4D EMT conservation for a
matter component with nonzero $\omega_r$. One can integrate this
equation to
\be
\rho_m =
\rho_{m0}a^{-3\frac{1+\omega}{1-\omega_r}},
\label{rho:evolution}
\ee where $\rho_{m0}$ is a constant. Using this in
(\ref{friedtr}) and taking $\Lambda_{eff}=0$ we can get our modified Friedman equation as
\be
H^2={\cal C}a^{-4}+\frac{1}{3M_p^2}\left(1-\omega_r\right)
\rho_{m0}a^{-3\frac{1+\omega}{1-\omega_r}},
\label{H2:evolution}
\ee
where $H=\frac{\dot{a}}{a}$, ${\cal C}$ is
a constant of integration and
$M_p^2=\frac{6 M_\ast^6}{T+T_r}$.
We see that an energy
component with an equation of state that has $\omega_r$ different
from zero acts as a fluid with an effective equation of state
given by \be \omega_\mathit{eff}=\frac{\omega + \omega_r}{1-\omega_r}.
\ee In particular we can get an accelerating universe
($\omega_\mathit{eff}<-1/3$) in the case of having just a matter
component as the only source of the Friedman equation having
$\omega =0$ if $\omega_r < -1/2$.

Formally, the extra-dimensional pressure, $p_r$, enters all the
equations as an extra contribution to the brane EMT with a
cosmological constant-like equation of state. So we would obtain
the same cosmological equations for an infinitesimally thin brane
if we considered an extra contribution to its energy-momentum
tensor like \be T_{M}^{N(br)} =
diag(\rho,-p,-p,-p)\delta(r)-\delta_M^Np_r\delta(r).\ee So taking
this redefined brane EMT one gets conventional conservation of
energy and momentum. In this case we could see our system (when
$p_r=-T_r + \omega_r \rho_m$) as a two fluid one where, besides an
extra contribution to the cosmological constant given by $T_r$,
there is an energy interchange between conventional matter and a
quintessence-like fluid that has pressure an energy density
proportional to the matter energy density, like in the asymptotic
solutions of models of interacting quintessence
\cite{Zimdahl:2001ar,Gasperini:2001pc}.

For the thick brane one can check that the time component of the
full 5D EMT conservation equation is indeed compatible with a time
dependent $p_r$ but only when the brane is thick and $\phi$ is
also time dependent, since its integration inside the brane
reads
\bea
\frac{1}{na^3 |_\epsilon}\int_{-\epsilon}^\epsilon
\left[\dot{T}_0^0+  3\frac{\dot{a}}{a}\left(T_0^0-T_x^{x}
\right)+\dot{\phi}\left(T_0^0-T_r^r\right)\right]n
a^3 e^{\phi} dr &
\nonumber \\
\simeq \dot{\rho}_m + 3H(1+w)\rho_m-\frac{1}{na^3
|_\epsilon}\int_{-\epsilon}^\epsilon
T_r^r \dot{\phi} n a^3 e^\phi dr =0,& \label{EMTcons2}\eea
where in the second line we have approximated $a(r,t)\simeq
a(\epsilon,t)$ for $|r|<\epsilon$. Comparing this equation with
eq.(\ref{EMTcons}) it is obvious that we need a time dependent
$\phi$ inside the brane (or, equivalently, a variable brane thickness) whenever we have matter with nonzero $\omega_r$. One can
also understand a nonzero $\omega_r$ as an extra coupling of
matter to the radion field, since after all $p_r$ is nothing but
the result of taking a variation of the brane Lagrangian with
respect to $e^{2\phi}$, the $rr$ component of the metric.

The obvious question is now: is it possible to have matter on the
brane such that its equation of state has $w_r$ different from
zero? We can argue that this is the case when these particles are
KK ``modes'' of matter localized on the brane. We use the quotes
because by KK modes we mean now modes such that part of its mass
comes from having a nontrivial profile $inside$ the brane. It is
easy to check that a KK mode of a scalar field in case of having a
flat circular extra dimension has $\omega_r =
m_{KK}^2/\left(m_{KK}^2 + m^2\right)$ where $m_{KK}$ is a mass
gained from a KK mechanism and $m$ is a ``bare'' 5D mass. However
we want to focus now on modes that are confined on the brane but
gain a mass from having a non-trivial profile in the
extra-dimension. The existence of such modes can be expected generically in topological defects like a domain wall \cite{Arodz:1993sy,Laine:2004ji}
or a vortex \cite{Goodband:1995rt}, and they
cannot be obtained in the thin limit since, as is no surprise,
they are non-analytical in the defect width. The modes studied in
\cite{Arodz:1993sy,Goodband:1995rt} are excitations of the fields
that create these topological defects and can also be understood
as oscillations of the width of the defects. So whenever there is
a field theory description of our brane one would expect that this kind of excitations, that imply oscillations of
the brane width, have a non-negligible (with respect to
the energy density) pressure along the extra dimension.

So we have seen that we can understand the cosmology of a brane universe
filled with a dust of particles that have pressure along the extra
dimension as a quintessence-like one. Now the relation of this
``effective quintessence fluid'' energy density and pressure to the
matter one can be easily explained, suggesting a solution to the
coincidence problem. But how could we understand this behavior
from the dynamics and interactions of the particles that
form this dust? As we have said, comparing eqs.(\ref{EMTcons}) and
(\ref{EMTcons2}) we see that we can atribute the apparent violation of
the EMT conservation equation for matter with non-zero $\omega_r$ to a
time dependent brane thickness, or radion field. Also using these
equations we can estimate the radion time dependence as $\dot{\phi}
\sim \omega_r \dot{\rho}_m /T_r \sim \omega_r H^3 M_p^2 /T_r$. It is
worth mentioning that one can
expect a conformal coupling for the radion in a 4D effective theory
(see for instance \cite{McFadden:2004se}), and a time dependent
conformally coupled scalar field would generate a time dependent mass
and an effective (anti)friction force for these particles in the 4D
theory (see also \cite{Casas:1991ky}). This would in turn produce an
apparent non-conservation (through an effective pressure) in the 4D EMT
\cite{Zimdahl:2000zm}, but the time dependence of the radion field
seems to be too small to attribute the non-standard features of the
cosmology to this effect. So a potential for the radion might be
relevant in an effective 4D description of this cosmology in terms of
interactions of these particles with a scalar field, and this
potential should have its origin in the physics generating the domain
wall. 

The estimation of the radion time dependence allows us to be a
bit more specific on size of the terms that we neglected in the
matching. Such terms are proportional to the integral inside the
brane of the 4D curvature or $\ddot{\phi}$, that are of order $\sim
H^2/M_b$ and $\sim
\omega_r H^4 M_p^2/(T_r M_b)$, respectively, and we have defined
$M_b\equiv 1/\epsilon$. Using now 
eqs.(\ref{rho:evolution},\ref{H2:evolution}) and comparing these terms
with the ones we kept in the matching we see that the term coming from
the integration of $\ddot{\phi}$ is negligible (unless $T_r$ is
extremelly small), while the term coming from the integration of the
4D curvature is negligible as long as 
\begin{equation}
\frac{T+T_r}{M_b} \ll M_\ast^3,\label{approximations}
\end{equation}
that we regard as our definition of a thin brane.

Now assuming that the Dark Matter particles have a large negative
value of $\omega_r$ one might build up a model to address the
coincidence problem. In this kind of model one would get
acceleration when the Dark Matter particles started dominating the
energy density of the universe. One should perform numerical
simulations in the effective 4D theory describing this braneworld to see how such a model fits the cosmological
observations, but recent studies show that models of interacting
quintessence can fit very well CMB~\cite{Olivares:2005tb} and
structure formation 
\cite{Amendola:2002kd} data (see also~\cite{Koivisto:2005nr}). The most recent supernova data might however get the simplest scenario
into trouble \cite{Riess:2004nr}, due to the need of a decelerated matter dominated phase at high redshifts.

To sum up: we have obtained the matching conditions for a thick
codimension one brane. Allowing for matter sources that have a
nonzero pressure component along the extra dimension we have seen
that our cosmological equations get modified in a way that allows
acceleration even if $\omega=0$. We have argued that particles having non-negligible pressure along the extra dimension can be associated with excitations of the domain wall impliying a variable brane thickness. In fact eq.(\ref{EMTcons2}) has the clear interpretation of relating the apparent non-conservation of the energy momentum-tensor for these particles to a time-dependent radion field inside the brane, or brane thickness. Our
results open the possibility of obtaining from the braneworld a
natural solution to the coincidence problem, very much along the
lines of the quintessence models of
\cite{Zimdahl:2001ar,Gasperini:2001pc} (in its asymptotic limit).
But in our case the time dependent scalar field is identified
with an extra dimensional component of the metric, instead of
being the string theory dilaton used in \cite{Gasperini:2001pc}.
It would be nice to obtain the 4D effective theory describing these particles from a
direct dimensional reduction of the braneworld taking
into account the radion field. Keeping track of the higher dimensional diffeomorphisms that can be associated with the conformal invariance of the 4D effective action like in \cite{McFadden:2004se} one might end up with a quintessence model along the lines of \cite{Zimdahl:2001ar,Gasperini:2001pc} or with a description in terms of effective forces of the type discussed in \cite{Zimdahl:2000zm}. Another necessary development
would be to find a higher dimensional Lagrangian with domain wall
solutions bounding stable particles with sizable and negative
$\omega_r$, realizing this mechanism. Such a model could provide the first natural microphysical realization of a
quintessence-like cosmology that addresses the coincidence problem. 

\begin{acknowledgments}
We would like to thank 
PPARC and DOE for financial support. I.N. would like to
thank Fermilab for hospitality during the completion of this work.
\end{acknowledgments}

\end{document}